\documentclass[conference]{IEEEtran}
\usepackage{graphicx}
\usepackage{amsmath}
\usepackage{amsfonts}
\usepackage{bm}
\usepackage{graphicx}
\usepackage{slashbox}
\usepackage{threeparttable}
\usepackage{color}
\usepackage{multirow}
\usepackage{pifont}
\usepackage{tabularx}
\usepackage[square, comma, sort&compress, numbers]{natbib}
\usepackage[caption=false]{subfig}
\usepackage{color}
\usepackage{comment}
\ifCLASSINFOpdf
\else
\fi
\begin{document}
	%
	\title{Towards a Uniform Architecture for the Efficient Implementation of 2D and 3D Deconvolutional Neural Networks on FPGAs }

	\author{\IEEEauthorblockN{Deguang Wang$^{1,2}$, Junzhong Shen$^{1,2}$, Mei Wen$^{1,2}$ and Chunyuan Zhang$^{1,2}$}
		\IEEEauthorblockA{$^{1}$College of Computer, $^{2}$National Key Laboratory for Parallel and Distributed Processing\\
			National University of Defense Technology,
			Changsha, China 410073\\
			Email: wangdeguang13@nudt.edu.cn}
	}
	
	
	%


	\maketitle
	
	\begin{abstract}
		Three-dimensional deconvolution is widely used in  many computer vision applications. 
		However, most previous works have only focused on accelerating 2D deconvolutional neural networks (DCNNs) on FPGAs, while the acceleration of 3D DCNNs has not been studied in depth as they have higher computational complexity and sparsity than 2D DCNNs. 
		In this paper, we focus on the acceleration of both 2D and 3D DCNNs on FPGAs by proposing efficient schemes for mapping 2D and 3D DCNNs on a uniform architecture. By implementing our design on the Xilinx VC709 platform for four real-life 2D and 3D DCNNs, we can achieve 
		up to 3.0 TOPS with high hardware efficiency. Comparisons with CPU and GPU solutions demonstrate that we can achieve an improvement of up to $63.3 \times$ in throughput relative to a CPU solution and an improvement of up to $8.3\times$ in energy efficiency compared to a GPU solution.
		
	\end{abstract}
	

	%
	\IEEEpeerreviewmaketitle

	\section{Introduction}
\begin{figure}[!t]
	\centering
	\includegraphics[width=2.7in]{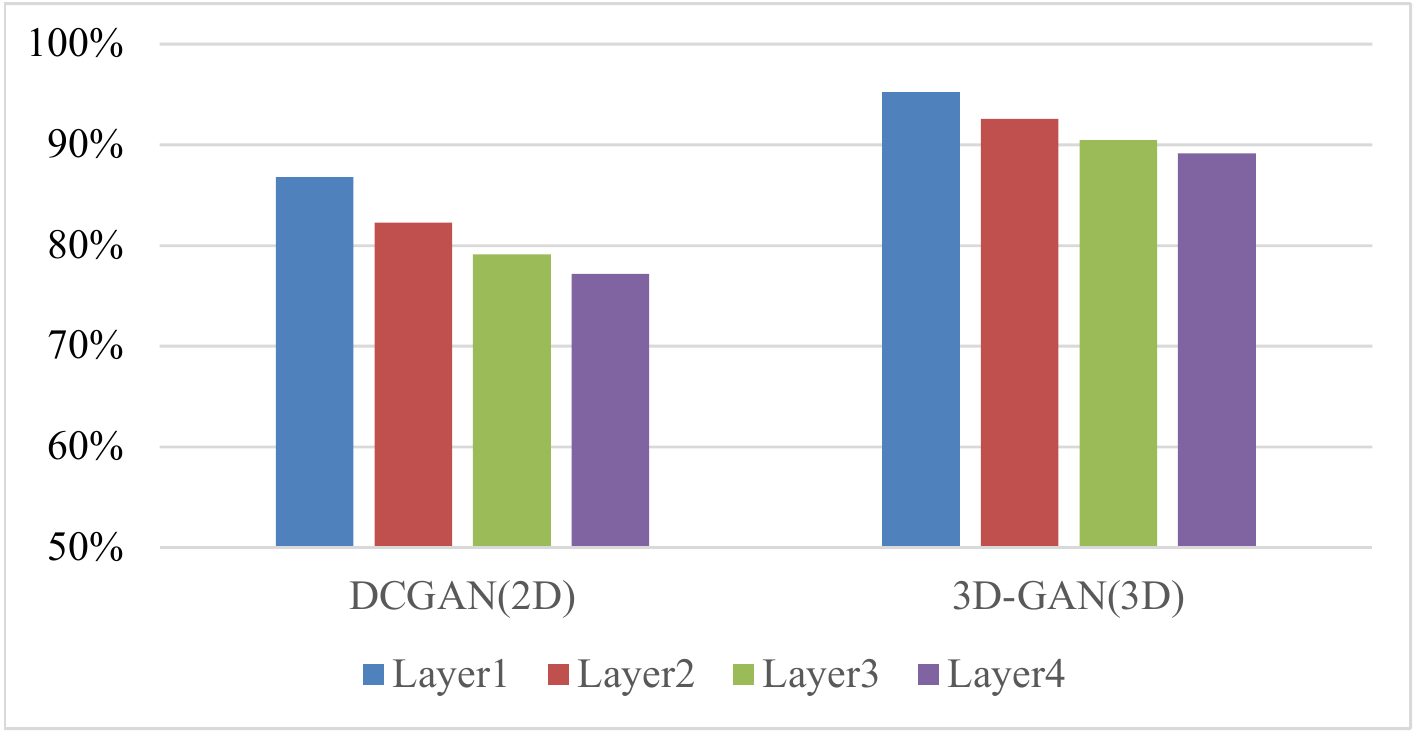}
	\vspace{-0.3em}
	\caption{Sparsity of the deconvolutional layers.}
	\label{Sparsity of deconvolution}
	\vspace{-1.6em}
	\end{figure}

	Recently, deconvolution has become widely used in the fields of computer vision, such as semantic segmentation \cite{long2015fully}, generative models \cite{radford2015unsupervised}, and high-resolution imaging \cite{dong2016accelerating}. 
	Because 3D images exist in most medical data used in clinical practice~\cite{milletari2016v}, 3D deconvolution has proven to be a better method than 2D deconvolution in some applications. Although the computational patterns of 2D and 3D deconvolutions are very similar, the computational complexity 
	and memory requirements of 3D deconvolution are much higher than in 2D deconvolution, making it challenging to design efficient accelerators for them. 
	In addition, deconvolution must insert `zero' into the input image 
	before implementing convolution operations, leading to the sparsity of the input image as well as the introduction of invalid operations (i.e., multiplications of zero). 
	According to our study, the sparsity of the input features of 3D deconvolution layers is higher than that of 2D deconvolution layers. As shown in Fig.~\ref{Sparsity of deconvolution}, the sparsity of the deconvolutional layers in an example of 3D deconvolutional neural networks (DCNNs) (i.e., 3D-GAN \cite{wu2016learning}) is clearly higher than for 2D DCNNs (i.e., DCGAN \cite{radford2015unsupervised}). Therefore, the sparsity contributes to the processing engine (PE) workload imbalance \cite{yan2018gna}. 
	
	Many studies \cite{zhang2015optimizing,qiu2016going,liu2017throughput} have primarily focused on accelerating convolutional neural networks (CNNs) on Field-Programmable Gate Arrays (FPGAs), due to the beneficial high performance and energy efficiency of FPGAs. 
	However, to the best of our knowledge, not much attention has been given to accelerate DCNNs, especially in 3D deconvolution. Given the similarity in the computational patterns of 2D and 3D deconvolutions, this work focuses on accelerating both of them on FPGA with a uniform architecture. 
	The contributions of this work are summarized as follows:
	\begin{enumerate}
		\item We propose a uniform architecture for efficient implementation of 2D and 3D DCNNs on FPGA.
		\item 	
		We propose a mapping scheme of 2D and 3D DCNNs on the uniform architecture, which can efficiently improve the parallel computational ability and computational efficiency of the accelerator.
		\item 	As a case study, we implement our design on an Xilinx VC709 board for four state-of-the-art 2D and 3D DCNNs: DCGAN, GP-GAN \cite{wu2017gp}, V-NET \cite{milletari2016v} and 3D-GAN. 
		Experimental results show that our implementation achieves an improvement of up to $63.3\times$ and $291.4\times$ in throughput and energy efficiency relative to CPU, and a $8.3\times$ energy efficiency gain over GPU.
	\end{enumerate}
	\vspace{-0.3em} 

	\section{Related Work}
	\begin{figure*}
		\centering
		\includegraphics[width=0.7\textwidth]{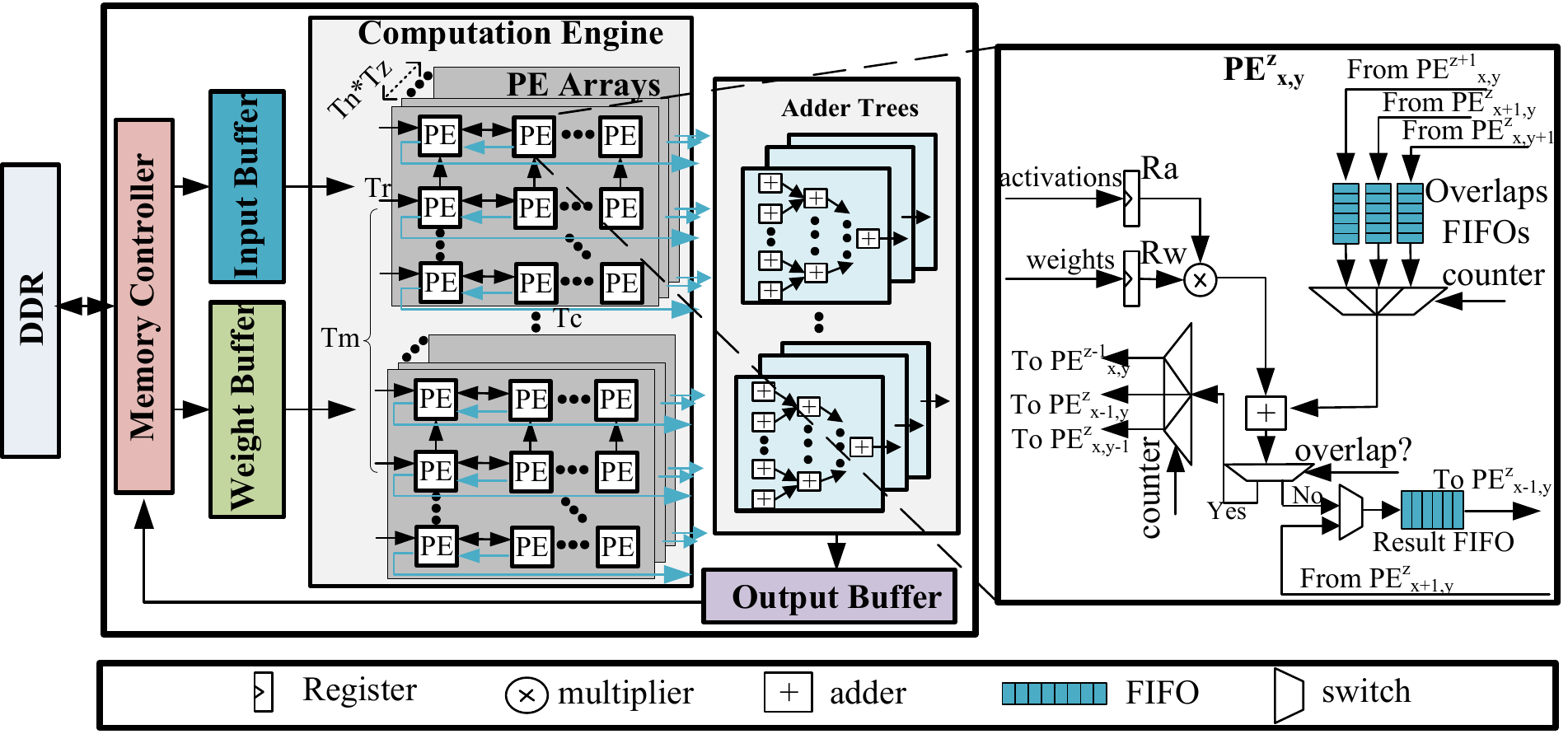}
		\vspace{-0.4em}
		\caption{An overview of our proposed architecture.}
		\label{architecture1}
		\vspace{-1.2em}
	\end{figure*}
	\begin{figure}[htbp]
	\centering
	\includegraphics[width=3.0in]{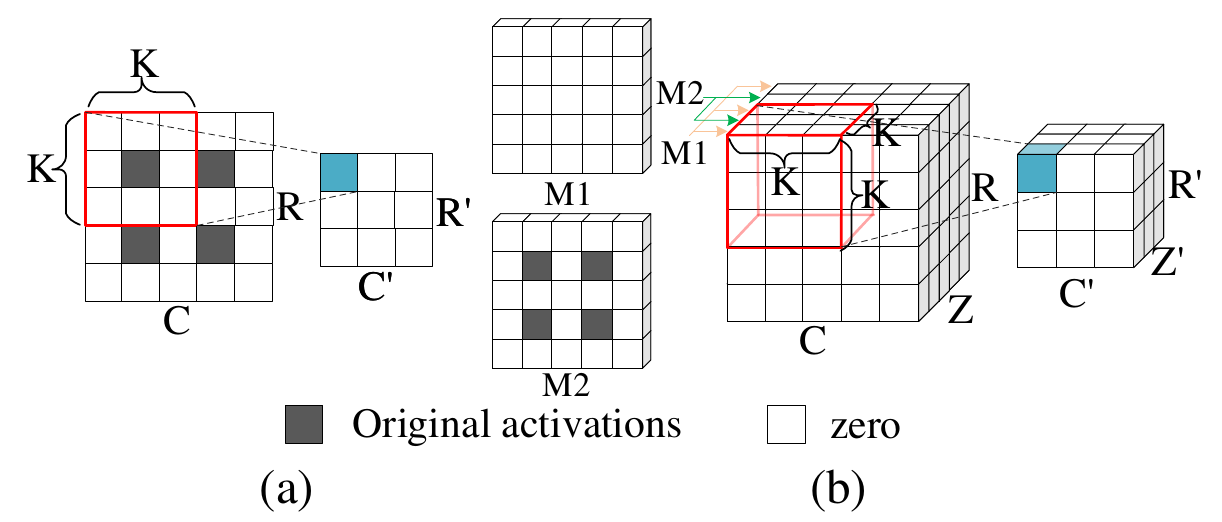}
	\vspace{-1.0em}
	\caption{Illustration of the process of 2D and 3D deconvolutions.}
	\label{comparasion_conv_deconv1}
	\vspace{-1.8em}
\end{figure}

	Few works have focused on accelerating deconvolutions \cite{yazdanbakhsh2018ganax, yazdanbakhsh2018flexigan, yan2018gna}. In \cite{yazdanbakhsh2018ganax, yazdanbakhsh2018flexigan}, the researchers addressed the accelerations of the deconvolution in generative adversarial networks(GANs). Yazdanbakhsh et al. \cite{yazdanbakhsh2018ganax} introduced a new architecture to alleviate the sources of inefficiency associated with the acceleration of GANs using conventional convolution accelerators by reorganizing the output computations. In \cite{yazdanbakhsh2018flexigan}, an end-to-end solution was devised to generate an optimized synthesizable FPGA accelerator from a high-level GAN specification, alleviating the challenges of inefficiency and resources underutilization faced by conventional convolutional accelerators. Yan et al. \cite{yan2018gna} proposed a novel mapping method called input oriented mapping (IOM, i.e., mapping each input computation task to each PE), which can efficiently overcome the inefficiency of PE computation. All the above mentioned works, however, only consider 2D DCNNs. To the best of our knowledge, we are the first to explore the acceleration of 2D and 3D DCNNs using a uniform architecture.

	\section{Background}

	Deconvolution is 
	similar to convolution operations, and the fundamental difference between them is that the original input feature maps of deconvolution requires the inseration of `zero' between the original input activations. 
	Fig.~\ref{comparasion_conv_deconv1} shows the process of 2D and 3D deconvolutions. 
	
	As Fig.~\ref{comparasion_conv_deconv1} (a) illustrates, for 2D deconvolution, the original input map is inserted with `zero' shown in white between the original input activations colored in gray. A $K\times K$ kernel then performs convolutions with the inserted feature map 
	to generate an $R'\times C'$ output map. Observed from Fig.~\ref{comparasion_conv_deconv1} (b), the process of 3D deconvolution is similar to that of 2D deconvolution. The original image is first inserted with `zero' between the  rows and columns of the 2D data tiles, which is identical to 2D deconvolution. In addition, it is also necessary to insert `zero' planes (i.e., the M1 plane) between every two 2D planes (i.e., the M2 plane) and a  $K\times K\times K$ kernel then performs convolutions with the inserted feature map 
	to generate an $R'\times C'\times Z'$ output map. 

	\begin{figure*}
	\centering
	\includegraphics[width=0.8\textwidth]{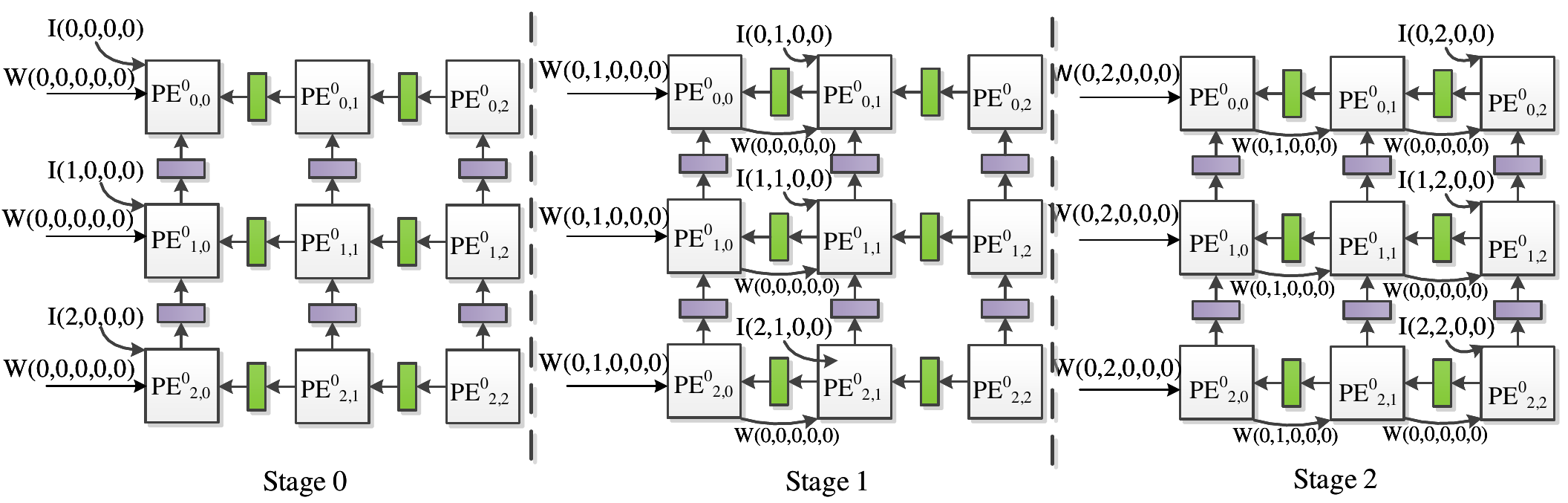}
	\vspace{-1.0em}
	\caption{Dataflow of the computation engine.}
	\label{dataflow}
	\vspace{-1.4em}
\end{figure*}	
\begin{figure}[!t]
	\centering
	\includegraphics[width=2.6in]{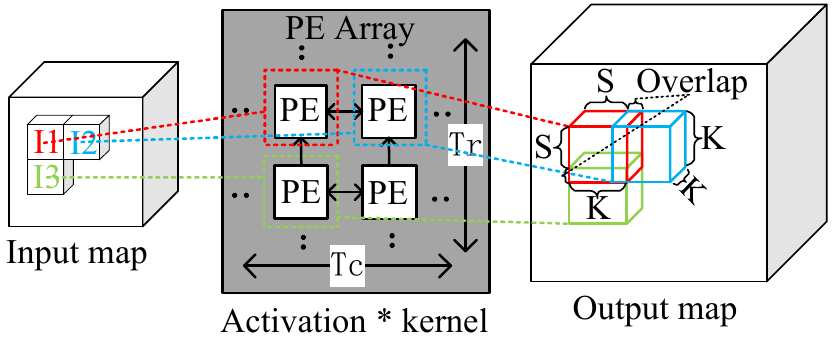}
	\vspace{-0.5em}
	\caption{Illustration of the 3D IOM method.}
	\label{input mapping method}
	\vspace{-1.8em}
\end{figure}
	\section{ The Proposed Architecture}
	\subsection{Architecture Overview}
	
	Fig.~\ref{architecture1} presents an overview of our proposed architecture for accelerating 2D and 3D deconvolutions. 
	The accelerator consists of a memory controller, three types of on-chip buffers, a kernel computation engine, and the adder trees. 
	Due to limited amount of on-chip memory of FPGAs, the source data and final results are stored in the off-chip 
	rate memory (i.e., the DDR).  
	The memory controller is used for fetching the input feature maps and weights from the DDR to the on-chip buffers, and storing the results into the DDR when they are available. In addition, one output feature map needs $N_c$ (i.e., input channels) input feature maps, due to the limited on-chip memory, it is difficult to cache all the input data needed for one feature map on chip. Hence, we use blocking to resolve this issue. We adopt three separate on-chip buffers to store input, output and weight blocks. 
	
	The computation engine is the most important component of our accelerator, which consists of a $T_m$ group of PEs. In each group, the PEs are organized as a 3D mesh architecture, which contains $T_n\times T_z$ 2D PE planes. In this work, we regard the PE plane as a PE array with $T_r\times T_c$ PEs. All PEs have direct connections to the input buffer, while only the leftmost PEs in each row have access to the weight buffer. The leftmost PEs in each row are responsible for collecting the results of the PEs of the same row and then deliver them to the following adder trees. The adder trees handle the additions of the results belonging to different feature maps. $T_m\times T_c\times T_z\times log_2 T_n$ adders are integrated in the adder trees to support a higher degree of parallelism.
	
	The architecture of the PE is presented in the right part of Fig.~\ref{architecture1}. It consists of two register files (i.e., \textit{Ra} and \textit{Rw}) to buffer the input activations and weights. In addition, three \textit{Overlap First-In-First-Outs (FIFOs)} (i.e., FIFO-Vs, FIFO-Hs and FIFO-Ds) are designed to deliver the overlaps in the results data from the adjacent PEs.
	The products yielded by the multipliers are conditionally added with the data from the \textit{Overlap FIFOs}. Once the current results are determined to be overlaps,  they will be sent to the \textit{Overlap FIFOs} of adjacent PEs, waiting to be added. Otherwise, they will be sent to the local \textit{Result FIFOs}. The results in the local FIFO of the current PE will be sent to the left PE once they have stored all the local results. 

	\subsection{3D IOM Method}


	Previous studies \cite{
	yazdanbakhsh2018ganax,yazdanbakhsh2018flexigan} have adopted the output oriented mapping (OOM, i.e., mapping each output computation task to each PE) for the computation of deconvolution layers. This method, however, does not eliminate invalid operations thereby resulting in low computational efficiency of the PEs. In \cite{yan2018gna}, Yan et al. proposed a novel mapping method called IOM, which can efficiently overcome the inefficiency of PE computation. Motivated by \cite{yan2018gna}, we propose a 3D version of IOM for the mapping of 3D deconvolution on the accelerator.
	
	Fig.~\ref{input mapping method} illustrates the 3D IOM method. I1$\sim$I3 are adjacent activations of the input map, and they are sent to three adjacent PEs of the PE array. In the PEs, each activation is multiplied by the $K\times K\times K$ kernel and generates a $K\times K\times K$ result block. The results are added to the corresponding location of the output maps. It is worth noting that some locations may overlap in the output maps and the overlapped elements of the same location should be added up to form the resulting output maps. 
	The overlap results from the PEs which are responsible for processing I2 and I3 are sent to the PE which is responsible for processing I1, and point-wise addition is performed. In each block, the length of the overlapping part is $K$-$S$. 
	
	In 3D deconvolution, the output feature map size is given by Eq. (1). Note that $I_H, I_W,  I_D, O_H, O_W,  O_D$ represent the height, width and depth of the input maps and output maps. However, at the edge of the output feature map, there is additional data padded. Thus, the padded data is removed from the final output feature map. The final result is equal to traditional convolution with `zero' inserted into the original input map. 
	\begin{equation}
	\begin{array}{l}
	O_{H} = (I_H -1) \times S+K. \\
	O_{W} = (I_W -1) \times S+K. \\
	O_{D} = (I_D -1) \times S+K.
	\end{array}	 
	\end{equation}

	We divide the dataflow in the PE arrays into three steps:
	
	\textbf{Loading activations and weights}: 
	Input blocks and weight blocks are firstly fetched into the input and weight buffers, activations and weights are loaded into the leftmost PEs in the 3D PE mesh from the input buffers and weight buffers. When the next column's PEs are empty, the next 
	group of activations are loaded into the next column's PEs in the next cycle. The activations in each PE are multiplied by all the weights of the corresponding kernels. The weights are also loaded into the leftmost PEs at the beginning of the process. When the weights are multiplied by activations in PEs, the weights are also sent to the next column's PEs. 
	The same column's PEs shares the weights. 
	
	\textbf{Computing}:
	After the activations and weights are loaded into the PEs, they are immediately sent to the multiplier  to yield the products in each PE. The results are then sent to the FIFOs. If the results overlap, they are send to Overlap FIFOs, else sent to Output FIFOs. Each PE performs $K\times K\times K$ multiplications to produce an output block. When th PEs process the overlapped part of the output blocks, the PEs load the overlapped elements from their FIFOs, and perform additions. 
	When the computation process in the direction of input channels (i.e., $T_n$) is complete, $T_n$ results are accumulated by the adder trees.
	
	\textbf{Writing Back}:
	When all the activations of the input blocks are complete and the overlaps are accumulated, the results, i.e., the output feature map is transferred to the output buffers. The results are accumulated until the input channels are complete, and the final outputs of output feature maps are then transferred to the external memory.        
	
	To explain this concept in more detail, we illustrate the dataflow of the PE arrays after applying the 3D IOM method on the architecture in Fig.~\ref{dataflow}. For the sake of simplicity, Fig.~\ref{dataflow} only shows the dataflow in a PE array, and the dataflow in other PE arrays are analogous. 
	Table \ref{definition} lists the definitions used in the explanation of the dataflow.
	At stage 0, the activations I(0,0,0,0)$\sim$I(2,0,0,0) and weights W(0,0,0,0,0) are sent separately to the leftmost columns of the PEs in the PE array, i.e. PE$^0_{0,0}$$\sim$PE$^0_{2,0}$, and they are multiplied.
	The overlaps produced from PE$^0_{1,0}$$\sim$PE$^0_{2,0}$ 
	is sent to their FIFO-Vs. 
	At stage 1, activations I(0,1,0,0)$\sim$I(2,1,0,0) are loaded into PE$^0_{0,1}$$\sim$PE$^0_{2,1}$. W(0,0,0,0,0) are moved to PE$^0_{0,1}$$\sim$PE$^0_{2,1}$, and they are then multiplied by the activations I(0,1,0,0)$\sim$I(2,1,0,0). Meanwhile, PE$^0_{0,0}$$\sim$PE$^0_{2,0}$ performs the multiplication by W(0,1,0,0,0). The overlaps produced by PE$^0_{0,1}$$\sim$PE$^0_{2,1}$ are then sent to their FIFO-Hs, and the overlaps produced by PE$^0_{1,0}$$\sim$PE$^0_{2,1}$ are sent to their FIFO-Vs. At stage 2, activations I(0,2,0,0)$\sim$I(2,2,0,0) are loaded into PE$^0_{0,2}$$\sim$PE$^0_{2,2}$. In the meantime, W(0,0,0,0,0) and W(0,1,0,0,0) are moved to PE$^0_{0,2}$$\sim$PE$^0_{2,2}$ and PE$^0_{0,1}$$\sim$PE$^0_{2,1}$, and then multiplied by the corresponding activations. The overlaps produced by PE$^0_{0,2}$$\sim$PE$^0_{2,2}$ are sent to their FIFO-Hs, and the overlaps produced by PE$^0_{1,0}$$\sim$PE$^0_{2,1}$ are sent to their FIFO-Vs. 
	\begin{table}[!t]
	\caption{Definition of the parameters.}
	\vspace{-0.8em}
	\label{definition}
	\begin{tabular}{l|l}
		\hline
		\parbox[l]{7mm}{\textbf{Parameter}} & \textbf{Description} \\ \hline
		I(i$_{h}$,i$_{w}$,i$_{d}$,i$_{c}$)   & {input activation from the i$_{c}$$^{th}$ input channel}   \\ \hline
		W(k$_{h}$,k$_{w}$,k$_{d}$,i$_{c}$,o$_{c}$)  & {weight from the {i$_{c}$}$^{th}$ channel of the o$_{c}$$^{th}$ filter}  \\ \hline
		W(k$_{h}$,k$_{w}$,k$_{d}$,i$_{c}$,o$_{c}$)  & {weight from the {i$_{c}$}$^{th}$ channel of the o$_{c}$$^{th}$ filter}  \\ \hline
	\end{tabular}
\vspace{-1.8em}
\end{table}	

	\subsection{Support for The Accelerations of 2D and 3D DCNNs}
	Our architecture is able to support the acceleration of both 2D and 3D DCNNs. For 3D DCNNs, $T_z$ PE arrays are used for the computations of an input feature map. In this way, $T_n\times T_z$ PE arrays can accelerate the computation of $T_n$ input feature maps simultaneously. For 2D DCNNs, we map the computations of an input feature map onto a PE array. Since the input feature maps are two-dimensional, we can use $T_n\times T_z$ PE arrays to compute $T_n\times T_z$ input feature maps in the meantime, while maintaining the size of the PE arrays (i.e., $T_r\times T_c$). In this case, the FIFO-D in each PE is disabled since there is no dataflow between adjacent PE arrays. Note that the dataflow in the PE arrays are identical when mapping 2D and 3D DCNNs on the computation engine. Since few control logics are required for supporting both 2D and 3D DCNNs in each PE, we omit the architecture details in Fig.~\ref{architecture1}. 

	\section{Experimental Results}
	As a case study, we evaluate our design using four representative DCNN models: DCGAN, GP-GAN, 3D-GAN and V-Net. All the deconvolutional layers of the selected DCNNs have uniform $3\times3$ and $3\times3\times3$ filters.

	We quantitatively compare our FPGA implementation of 2D and 3D DCNNs with two other platforms: (1) a ten-core Intel E5 CPU (2.8 GHz) and (2) a NVIDIA GeForce GTX 1080 GPU. Our accelerator design is implemented on the Xilinx VC709 clocked at 200MHz, which contains a Virtex-7 690t FPGA and two 4GB DDR3 DRAMs. 
	
	
	Table \ref{configuration} illustrates the configuration of the parameters of our benchmarks. 
	Note that we use 16-bit fixed activations and weights for all the benchmarks in our experiment. To avoid the reconfiguration overhead, we use an accelerator with fixed configurations for all the benchmarks. We use $T_m\times T_n\times T_z\times T_r\times T_c = 2,048$ PEs in total. 
	
	Table \ref{resource} reports the resource utilization of our accelerator. The Digital Signal Processors (DSPs) and Look-up Tables (LUTs) dominate the resource consumption, and mainly utilized for implementing multipliers and adders, respectively. 
		\begin{table}[!t]
	\centering
	\caption{Configurations of the computation engine.}
	\vspace{-0.8em}
	\label{configuration}
	\begin{tabular}{l|c|c|c|c|c|c}
		\hline
		\textbf{Benchmarks} & \textbf{$T_m$} & \textbf{$T_n$} &   \textbf{$T_z$} &   \textbf{$T_r$} &   \textbf{$T_c$} &   \textbf{data width} \\ \hline 
		2D DCNNs   &2  &64   &1 &4 &4 &16   \\ \hline
		3D DCNNs  &2  &16  &4  &4  &4 &16 \\ \hline
	\end{tabular}
\vspace{-1.8em}
\end{table}
	\begin{table}[!t]
		\centering
		\caption{Resource utilization of Xilinx VC709.}
		\vspace{-0.8em}
		\label{resource}
		\begin{tabular}{l|c|c|c|c}
			\hline
			\textbf{Resource} & \textbf{DSP48Es} & \textbf{BRAMs} &   \textbf{Flip-Flops} & \textbf{LUTs} \\ \hline 
			Utilization   &2304  &712   &566182  & 292292  \\ \hline
			percentage(\%)  &64.00  &48.44  &65.34   & 67.48  \\ \hline
		\end{tabular}
	\vspace{-1.5em}
	\end{table}
\begin{figure}[!t]
	\centering
	\subfloat[]{\includegraphics[width=1.8in]{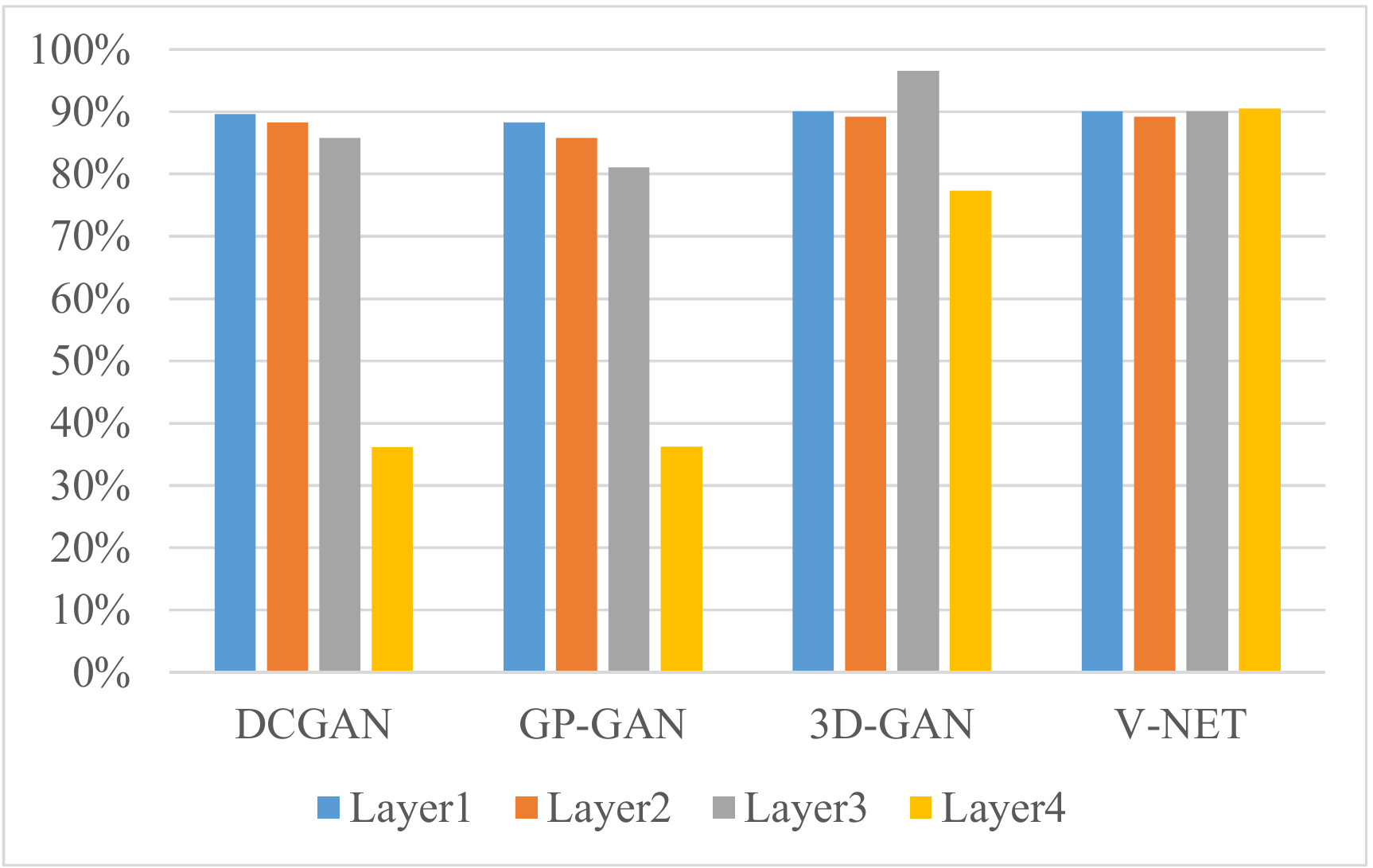}} 
	\subfloat[]{\includegraphics[width=1.8in]{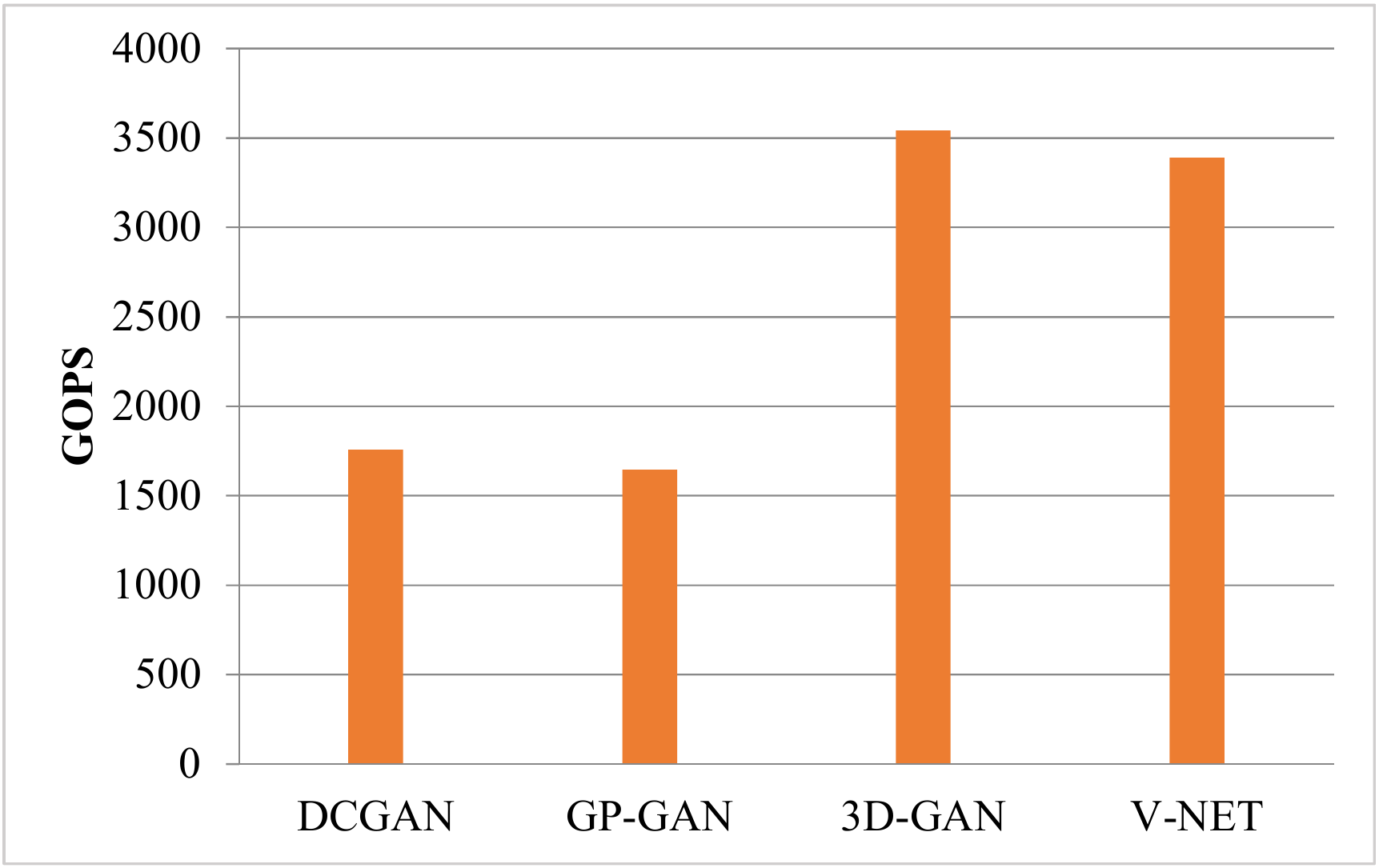}}
	\vspace{-0.5em}
	\caption{Results for 2D and 3D DCNNs: (a) the PE utilization; (b) throughput.} 
	\label{performance} 
	\vspace{-2.0em}
\end{figure}
\begin{figure}[!t]
	\centering
	\subfloat[]{\includegraphics[width=1.8in]{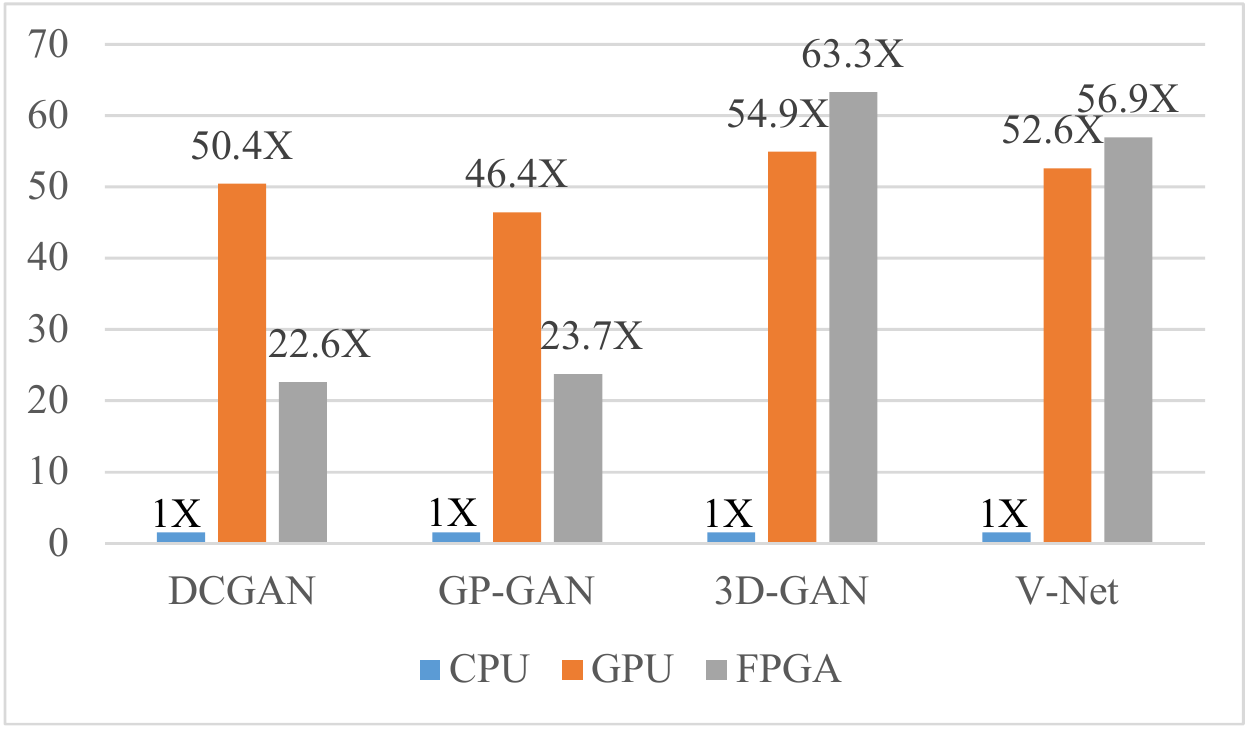}}
	\subfloat[]{\includegraphics[width=1.8in]{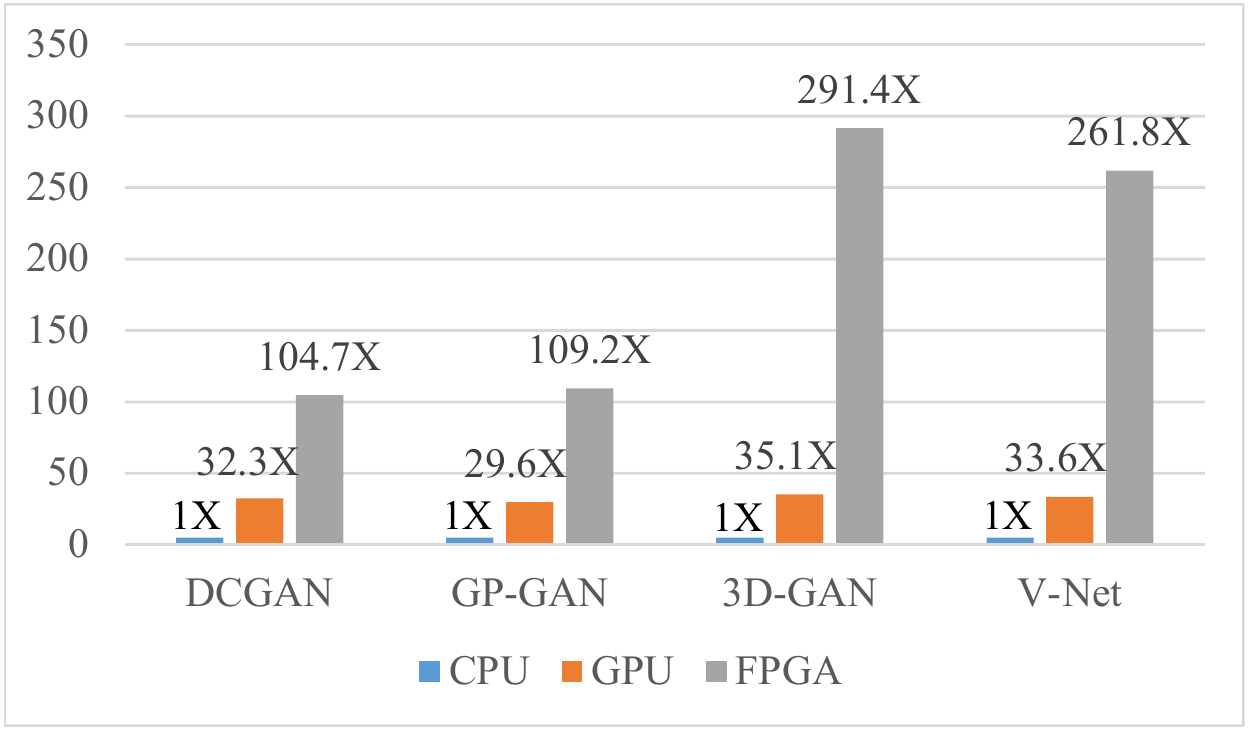}}
	\vspace{-0.5em}
	\caption{Comparisons of CPU, GPU and FPGA solutions: (a) relative performance; (b) relative energy efficiency.} 
	\label{speedup} 
	\vspace{-2.0em}
\end{figure} 
	Fig.~\ref{performance} presents PE utilization about the accelerator. Note that the PE utilization is defined as the ratio of the computation time occupied in total time.  For all benchmarks, our accelerator can achieve up over 90\% of PE utilization. It demonstrates the effectiveness of our mapping and the uniform architecture for 2D and 3D deconvolutions. Note that the fourth layers of DCGAN and GP-GAN are bottlenecked by the memory access, which results in a reduction of PE utilization. In addition, it is also clear from Fig.~\ref{performance} that we can achieve state-of-the-art performance (1.5TOPS$\sim$3.0TOPS) for all the benchmarks. 
	Because the higher sparsity of 3D deconvolution and the large amount of data delivered between PEs, the performance of 3D deconvolution on FPGA outperforms that of 2D deconvolution.  
	
	For the benchmarks in our experiment, 
	we compare our work with CPU and GPU, as shown in Fig.~\ref{speedup}. The performance of our method on the accelerator outperforms that of the CPU by $22.7\times$ to $63.3\times$. 
	Concerning energy efficiency, our method outperforms the CPU by $104.7\times$ to $291.4\times$ 
	and outperforms the GPU by $3.3\times$ to $8.3\times$.

	\section{Conclusion}
	In this paper, we proposed a 2D and 3D deconvolution accelerator based on a uniform architecture on FPGA. We employed a mapping scheme of 2D and 3D deconvolutions on this architecture. 
	To the best of our knowledge, this is the first work to implement 2D and 3D DCNNs on FPGA. By exploring the data transference between adjacent PEs without invalid operations, our design achieves an acceleration of $63.3 \times$ compared with CPU implementation, and an energy efficiency improvement of $8.3\times$ compared with designs running on a GTX 1080 GPU. 
	\ifCLASSOPTIONcaptionsoff
	\newpage
	\fi

	\footnotesize
	\bibliographystyle{IEEEtran}
	\bibliography{IEEEabrv,./document.bbl}

\begin{thebibliography}{10}
\providecommand{\url}[1]{#1}
\csname url@samestyle\endcsname
\providecommand{\newblock}{\relax}
\providecommand{\bibinfo}[2]{#2}
\providecommand{\BIBentrySTDinterwordspacing}{\spaceskip=0pt\relax}
\providecommand{\BIBentryALTinterwordstretchfactor}{4}
\providecommand{\BIBentryALTinterwordspacing}{\spaceskip=\fontdimen2\font plus
\BIBentryALTinterwordstretchfactor\fontdimen3\font minus
  \fontdimen4\font\relax}
\providecommand{\BIBforeignlanguage}[2]{{%
\expandafter\ifx\csname l@#1\endcsname\relax
\typeout{** WARNING: IEEEtran.bst: No hyphenation pattern has been}%
\typeout{** loaded for the language `#1'. Using the pattern for}%
\typeout{** the default language instead.}%
\else
\language=\csname l@#1\endcsname
\fi
#2}}
\providecommand{\BIBdecl}{\relax}
\BIBdecl

\bibitem{long2015fully}
J.~Long, E.~Shelhamer, and T.~Darrell, ``Fully convolutional networks for
  semantic segmentation,'' in \emph{Proceedings of the IEEE conference on
  computer vision and pattern recognition}, 2015, pp. 3431--3440.

\bibitem{radford2015unsupervised}
A.~Radford, L.~Metz, and S.~Chintala, ``Unsupervised representation learning
  with deep convolutional generative adversarial networks,'' \emph{arXiv
  preprint arXiv:1511.06434}, 2015.

\bibitem{dong2016accelerating}
C.~Dong, C.~C. Loy, and X.~Tang, ``Accelerating the super-resolution
  convolutional neural network,'' in \emph{European Conference on Computer
  Vision}.\hskip 1em plus 0.5em minus 0.4em\relax Springer, 2016, pp. 391--407.

\bibitem{milletari2016v}
F.~Milletari, N.~Navab, and S.-A. Ahmadi, ``V-net: Fully convolutional neural
  networks for volumetric medical image segmentation,'' in \emph{3D Vision
  (3DV), 2016 Fourth International Conference on}.\hskip 1em plus 0.5em minus
  0.4em\relax IEEE, 2016, pp. 565--571.

\bibitem{wu2016learning}
J.~Wu, C.~Zhang, T.~Xue, B.~Freeman, and J.~Tenenbaum, ``Learning a
  probabilistic latent space of object shapes via 3d generative-adversarial
  modeling,'' in \emph{Advances in Neural Information Processing Systems},
  2016, pp. 82--90.

\bibitem{yan2018gna}
J.~Yan, S.~Yin, F.~Tu, L.~Liu, and S.~Wei, ``Gna: Reconfigurable and efficient
  architecture for generative network acceleration,'' \emph{IEEE Transactions
  on Computer-Aided Design of Integrated Circuits and Systems}, 2018.

\bibitem{zhang2015optimizing}
C.~Zhang, P.~Li, G.~Sun, Y.~Guan, B.~Xiao, and J.~Cong, ``Optimizing fpga-based
  accelerator design for deep convolutional neural networks,'' in
  \emph{Proceedings of the 2015 ACM/SIGDA International Symposium on
  Field-Programmable Gate Arrays}.\hskip 1em plus 0.5em minus 0.4em\relax ACM,
  2015, pp. 161--170.

\bibitem{qiu2016going}
J.~Qiu, J.~Wang, S.~Yao, K.~Guo, B.~Li, E.~Zhou, J.~Yu, T.~Tang, N.~Xu, S.~Song
  \emph{et~al.}, ``Going deeper with embedded fpga platform for convolutional
  neural network,'' in \emph{Proceedings of the 2016 ACM/SIGDA International
  Symposium on Field-Programmable Gate Arrays}.\hskip 1em plus 0.5em minus
  0.4em\relax ACM, 2016, pp. 26--35.

\bibitem{liu2017throughput}
Z.~Liu, Y.~Dou, J.~Jiang, J.~Xu, S.~Li, Y.~Zhou, and Y.~Xu,
  ``Throughput-optimized fpga accelerator for deep convolutional neural
  networks,'' \emph{ACM Transactions on Reconfigurable Technology and Systems
  (TRETS)}, vol.~10, no.~3, p.~17, 2017.

\bibitem{wu2017gp}
H.~Wu, S.~Zheng, J.~Zhang, and K.~Huang, ``Gp-gan: Towards realistic
  high-resolution image blending,'' \emph{arXiv preprint arXiv:1703.07195},
  2017.

\bibitem{yazdanbakhsh2018ganax}
A.~Yazdanbakhsh, H.~Falahati, P.~J. Wolfe, K.~Samadi, N.~S. Kim, and
  H.~Esmaeilzadeh, ``Ganax: A unified mimd-simd acceleration for generative
  adversarial networks,'' \emph{arXiv preprint arXiv:1806.01107}, 2018.

\bibitem{yazdanbakhsh2018flexigan}
A.~Yazdanbakhsh, M.~Brzozowski, B.~Khaleghi, S.~Ghodrati, K.~Samadi, N.~S. Kim,
  and H.~Esmaeilzadeh, ``Flexigan: An end-to-end solution for fpga acceleration
  of generative adversarial networks,'' in \emph{2018 IEEE 26th Annual
  International Symposium on Field-Programmable Custom Computing Machines
  (FCCM)}.\hskip 1em plus 0.5em minus 0.4em\relax IEEE, 2018, pp. 65--72.

\end{thebibliography}
\end{document}